\documentstyle[psfig]{mn}

\newif\ifAMStwofonts
\AMStwofontstrue

\def\gtrsim{\mathrel{\hbox{\rlap{\hbox{\lower4pt\hbox{$\sim$}}}\hbox{$>$}}}}

\ifoldfss
  \ifCUPmtlplainloaded \else
    \NewTextAlphabet{textbfit} {cmbxti10} {}
    \NewTextAlphabet{textbfss} {cmssbx10} {}
    \NewMathAlphabet{mathbfit} {cmbxti10} {} 
    \NewMathAlphabet{mathbfss} {cmssbx10} {} 
  \fi
  \ifAMStwofonts
    \ifCUPmtlplainloaded \else
      \NewSymbolFont{upmath} {eurm10}
      \NewSymbolFont{AMSa} {msam10}
      \NewMathSymbol{\upi}     {0}{upmath}{19}
      \NewMathSymbol{\umu}     {0}{upmath}{16}
      \NewMathSymbol{\upartial}{0}{upmath}{40}
      \NewMathSymbol{\leqslant}{3}{AMSa}{36}
      \NewMathSymbol{\geqslant}{3}{AMSa}{3E}

    \fi
  \fi
\fi 

\ifnfssone
  \newmathalphabet{\mathit}
  \addtoversion{normal}{\mathit}{cmr}{m}{it}
  \addtoversion{bold}{\mathit}{cmr}{bx}{it}
  \newmathalphabet{\mathbfit} 
  \addtoversion{normal}{\mathbfit}{cmr}{bx}{it}
  \addtoversion{bold}{\mathbfit}{cmr}{bx}{it}
  \newmathalphabet{\mathbfss} 
  \addtoversion{normal}{\mathbfss}{cmss}{bx}{n}
  \addtoversion{bold}{\mathbfss}{cmss}{bx}{n}
  \ifAMStwofonts
    \ifCUPmtlplainloaded \else
      \UseAMStwoboldmath
      \makeatletter
      \new@mathgroup\upmath@group
      \define@mathgroup\mv@normal\upmath@group{eur}{m}{n}
      \define@mathgroup\mv@bold\upmath@group{eur}{b}{n}
      \edef\UPM{\hexnumber\upmath@group}
      \new@mathgroup\amsa@group
      \define@mathgroup\mv@normal\amsa@group{msa}{m}{n}
      \define@mathgroup\mv@bold\amsa@group{msa}{m}{n}
      \edef\AMSa{\hexnumber\amsa@group}
      \makeatother
      \mathchardef\upi="0\UPM19
      \mathchardef\umu="0\UPM16
      \mathchardef\upartial="0\UPM40
      \mathchardef\leqslant="3\AMSa36
      \mathchardef\geqslant="3\AMSa3E
    \fi
  \fi
\fi 

\ifnfsstwo
  \DeclareMathAlphabet{\mathbfit}{OT1}{cmr}{bx}{it}
  \SetMathAlphabet\mathbfit{bold}{OT1}{cmr}{bx}{it}
  \DeclareMathAlphabet{\mathbfss}{OT1}{cmss}{bx}{n}
  \SetMathAlphabet\mathbfss{bold}{OT1}{cmss}{bx}{n}
  \ifAMStwofonts
    \ifCUPmtlplainloaded \else
      \DeclareSymbolFont{UPM}{U}{eur}{m}{n}
      \SetSymbolFont{UPM}{bold}{U}{eur}{b}{n}
      \DeclareSymbolFont{AMSa}{U}{msa}{m}{n}
      \DeclareMathSymbol{\upi}{0}{UPM}{"19}
      \DeclareMathSymbol{\umu}{0}{UPM}{"16}
      \DeclareMathSymbol{\upartial}{0}{UPM}{"40}
      \DeclareMathSymbol{\leqslant}{3}{AMSa}{"36}
      \DeclareMathSymbol{\geqslant}{3}{AMSa}{"3E}
    \fi
  \fi
\fi 

\ifCUPmtlplainloaded \else
  \ifAMStwofonts \else 
    \def\upi{\pi}
    \def\umu{\mu}
    \def\upartial{\partial}
  \fi
\fi

\title{Error analysis of the photometric redshift tecnique}
\author[A. Fern\'andez-Soto et al.]
       {A.~Fern\'andez-Soto,$^1$\thanks{Marie Curie Fellow.}\thanks{Email:
        fsoto@merate.mi.astro.it} K.M.~Lanzetta,$^2$ H.-W.~Chen,$^3$ 
        B.~Levine,$^2$ N.~Yahata$^2$ \\
        $^1$Osservatorio Astronomico di Brera, Via Bianchi 46, Merate
        (LC), I-23807, Italy\\
        $^2$Department of Physics and Astronomy, State University of New 
        York at Stony Brook, Stony Brook, NY 11794-3800, U.S.A.\\
        $^3$The Observatories of the Carnegie Institution of Washington,
        813 Santa Barbara Street, Pasadena, CA 91101, U.S.A.}

\date{Accepted ---.
      Received ---;
      in original form ---}

\pagerange{\pageref{firstpage}--\pageref{lastpage}}
\pubyear{2001}

\begin{document}
\maketitle

\label{firstpage}

\begin{abstract}
We present a calculation of the systematic component of the error budget in
the photometric redshift technique. We make use of it to describe a simple
technique that allows for the assignation of confidence limits to redshift
measurements obtained through photometric methods. We show that our
technique, through the calculation of a redshift probability function, gives
complete information on the probable redshift of an object and its associated
confidence intervals. This information can and must be used in the
calculation of any observable quantity which makes use of the redshift.
\end{abstract}


\section{Introduction}
One of the most basic operations that need to be performed in Cosmology is
the measurement of the redshift of any given object. As is well known, such
an operation can be more or less readily fulfilled depending on several
factors like the brightness of the object, the available instrumentation, and
the analysis technique. Given that advancement in science is always driven by
work at the very edge of feasibility, it is not surprising that cosmologists
often find themselves trying to measure redshifts from objects that are too
faint for even the most advanced spectrographs.

This dearth of photons forces astronomers to try and identify spectral
features in noisy spectra. More often than is usually believed this causes
mistakes that are not easily noticed. This is because spectroscopic redshift
errors are often due to human biases (as when the observer makes a choice to
identify a possible emission/absorption line amongst comparable noise peaks;
or to assign line identifications based on his/her previous experience or
personal preference), and also because often the spectroscopic information is
not made available for general scrutiny. The particular case of the Hubble
Deep Field \cite{williams96}, without doubt the most deeply observed patch of
the sky, is palmary: not less than five papers have been published with
spectroscopic redshifts of {\it multiple, different} HDF objects that have
later been retracted as `erroneous'; see Fern\'andez-Soto et al. (2001, FS01
from here on) for a complete analysis. The effect of those errors in
subsequent papers that made use of the wrong spectroscopic lists is difficult
to ascertain. It is not straightforward to find a solution to this problem
within the usual techniques, because of the sources of bias listed above.

The use of photometric techniques to measure redshifts was suggested as early
as 1962 by Baum \cite{baum62}, and other authors (Koo 1981, Butchins 1981,
Loh \& Spillar 1986) pioneered similar ideas to overcome the difficulties
associated to spectroscopy of very faint sources. Photometric redshift
techniques boomed in the mid 1990s with the arrival of the Hubble Deep
Fields--extremely deep images in which exquisite photometry could be
performed on thousands of galaxies, over 90 per cent of which were too faint
for any spectrograph available at the time, nowadays, or in the near
future. Several groups have perfected different approaches (for example
Lanzetta, Yahil, \& Fern\'andez-Soto 1996; Gwyn \& Hartwick 1996; Sawicki,
Lin, \& Yee 1997; Wang, Bahcall, \& Turner 1998; Fern\'andez-Soto, Lanzetta,
\& Yahil 1999--hereafter FLY99, Furusawa et al. 2000; Ben\'{\i}tez 2000;
Yahata et al. 2000--hereafter Y00; Fontana et al. 2000; Massarotti et
al. 2001) and nowadays it can be said that photometric redshift techniques
are an integral part of the standard cosmological toolbox.

Most cosmologists will concur with the opinion that the photometric
techniques are useful because they expand the volume of `distance-luminosity'
space where redshifts can be measured--even if the values so measured are
{\it somehow `less precise'} than the spectroscopic ones, a payback most are
willing to accept. We find this very concept (the {\it `lack of precision'})
very difficult to evaluate. It is uncomfortable for any scientist to talk
about the accuracy of a measurement whenever a confidence interval has not
been assigned to it, and as has been exposed above, that is precisely the
problem with spectroscopy of faint sources.

In a previous paper (FS01) we showed that our particular technique is able to
measure redshifts of faint objects with a reliability which is comparable (if
not superior) to that of the traditional spectroscopic method. We present in
this work a simple method that allows for the calculation of accurate
confidence intervals around photometric redshift measurements. The use of
these confidence intervals should solve the problem of the so-called {\it
`catastrophic errors'}, when the photometric technique gives results that are
very different from the spectroscopic ones. We intend to show that in those
apparently discordant cases, either the values are in fact consistent (when
the photometric value is actually compatible with the spectroscopic one
within an acceptable probability level) or the problem is serious enough to
call for a revision of {\it both} values--when they are incompatible to a
large degree of confidence.

We further suggest that the photometric redshifts together with their
associated probability functions, {\it can and must} be used in the
calculation of any quantity which is derived from the redshifts, in order to
perform an adequate error assessment of the results.

The structure of this paper is as follows: in Section \ref{seccat} we present
the catalogues of photometric/spectroscopic redshifts over which our
technique is tested. Section \ref{secerr} contains the description and
measurement of the sources of error (photometric and systematic) present in
the photometric redshift determination. We present and apply the technique to
estimate errors in Section \ref{sectec}, and discuss the results in Section
\ref{secdis}. Our conclusions are resumed in Section \ref{seccon}

\section{Photometric and spectroscopic data}
\label{seccat}

We will use the catalogue presented by FS01 (which in turn is based on the
spectroscopic catalogue by Cohen et al. 2000, C00 hereafter), as a basis to
calibrate the errors in our photometric measurements. The photometric data
used in the analysis include space images (Hubble Space Telescope optical
observations through the filters F300W, F450W, F606W and F814W), and
ground-based observations taken at Kitt Peak in the $J$, $H$, and $K$
bands. A few changes have been done, as follows:

\begin{enumerate}
\item Three of the spectroscopic redshifts that were discussed as possibly 
wrong by FS01 on the basis of the photometric information have been retracted
by Cohen (2001, C01 hereafter)--we use the new values. 

\item Another object under discussion has been remeasured by Dawson et
al. (2001, D01 hereafter). It is HDF36414\_1143, $z_{\rm sp}=1.524$, and has
been found to be in better agreement with our value ($z_{\rm ph}=1.32$) than
with that listed in C00 ($z=0.548$). We adopt the new spectroscopic value.

\item The rest of discrepant objects presented in FS01 are used with the same
considerations there presented. In particular, objects marked as `uncertain'
are not used in the calculations that follow. We note though that Massarotti
et al. (2001), also based on photometric considerations, disagree with us
in two of the objects.

\item One new object (HDF36453\_1143, $R=24.00$, $z_{\rm sp}=0.485$) was
added by C01 to the spectroscopic sample. It corresponds to object \#81 in
our catalogue, with $z_{\rm ph}=0.64$. Another object added in C01
(HDF36377\_1235) does not lie in the area studied by us.

\item An extra ten new objects from D01 are included in the sample. They are
listed in Table 1, together with HDF36414\_1143 (discussed above). 
\end{enumerate}

\begin{table*}
 \centering
  \begin{minipage}{140mm}
  \caption{New spectroscopic redshifts presented by D01 and added to our 
sample. The magnitude $AB(8140)$ comes from D01. The value of $z_{\rm ph}$ 
is from our new catalogue, still unpublished, which includes NICMOS 
observations.}
  \begin{tabular}{l c c c c c c c}
Object & RA(2000) & Dec(2000) & $AB(8140)$ & $z_{\rm sp}$ & \#(FLY99) & $z_{\rm ph}$ \\
 HDF36459\_1326 & 12:36:45.855 & 62:13:25.81 & 24.07 & 0.847 &  890 & 0.75\\
 HDF36485\_1317 & 12:36:48.474 & 62:13:16.62 & 23.45 & 0.474 &  775 & 0.27\\
 HDF36498\_1419 & 12:36:49.804 & 62:14:19.15 & 25.59 & 0.425 & 1035 & 2.26\\
 HDF36548\_1258 & 12:36:54.805 & 62:12:58.05 & 24.45 & 0.851 &  512 & 0.94\\
 HDF36582\_1307 & 12:36:58.190 & 62:13:06.58 & 24.57 & 0.475 &  496 & 0.53\\
 HDF36494\_1215 & 12:36:49.365 & 62:12:14.64 & 24.91 & 0.934 &  274 & 0.97\\
 HDF36478\_1218 & 12:36:47.838 & 62:12:18.30 & 28.26 & 0.102 &  --- &  ---\\
 HDF36438\_1252 & 12:36:43.822 & 62:12:51.96 & 24.96 & 1.013 &  735 & 0.91\\
 HDF36433\_1239 & 12:36:43.253 & 62:12:38.86 & 24.86 & 2.442 &  664 & 2.46\\
 HDF36447\_1144 & 12:36:44.734 & 62:11:43.77 & 24.77 & 0.558 &  108 & 0.67
\footnote{This object was misidentified by D01 with object 105 in our 
catalogue, and dubbed a `catastrophic error' of the photometric
technique. Object 108 is by far a better fit both to the position and the
magnitude of their source than object 105. We did point this to the
authors--together with other misidentifications in their paper--but they
somehow neglected to correct it in the final published version}\\
 HDF36423\_1126 & 12:36:42.284 & 62:11:26.18 & 25.09 & 0.559 &   14 & 0.64\\
 HDF36414\_1143 & 12:36:41.427 & 62:11:42.89 & 24.99 & 1.524 &  200 & 1.32
\end{tabular}
\end{minipage}
\end{table*}

The total list of photometric/spectroscopic redshift pairs is now composed of
153 values.

\section{Sources of error in the photometric redshift technique}
\label{secerr}

\subsection{The two sources of error}

As was presented in previous papers (Lanzetta, Fern\'andez-Soto \& Yahil
1998--hereafter LFY98, FLY99, Y00, FS01), the sources of error in the
photometric redshift measurements are twofold. There is an obvious
uncertainty in the redshift which is associated to the uncertainty in the
photometric measurements, and this is taken into account in our calculation
of the redshift likelihood function:

\begin{equation}
L(z,T)=\prod_{i=1}^{N} \exp\left\{-\frac{1}{2}\left[
\frac{f_i-AF_i(z,T)}{\sigma_i} \right] ^2 \right\}
\end{equation}

\noindent where the product extends to the number of observed filters, $A$ is
a normalization constant, $f_i$ and $\sigma_i$ are the flux and associated
error of the source measured in the $i$-th band, and $F_i(z,T)$ are the model
fluxes for a galaxy of type $T$ at redshift $z$ in the $i$-th band.

In principle the likelihood function determined this way should allow us to
calculate confidence limits in the parameters of interest (in our case the
errors associated to $z$). But this only applies to the cases in which {\it
the fitted model represents a good fit to the data}. This is not the case of
our technique. The reason for this is that the discrete number of templates
used to produce the model fluxes (six in our case) cannot be realistically
expected to reproduce the spectral energy distributions of all galaxies. This
fact will be particularly notorious for bright galaxies, where the
high-quality photometry will amplify any difference between the model and
the observations, hence producing two effects:

\begin{enumerate}
\item A very bad (in $\chi2$ terms) fit, even for a perfectly well-determined 
photometric redshift, whenever the source is bright, and

\item An extra dispersion in the values of $z_{\rm ph}$, that we will refer
to as {\it cosmic variance}, {\it SED variance}, or {\it systematic dispersion}.
\end{enumerate}

The effect of cosmic variance dominates the error budget for all bright
sources (when the photometric error is small enough to allow the
`imperfections' of the model SED to be noticed), whereas it is negligible for
faint sources.

We try in this section to overcome the problem posed by the first item above
by means of measuring the effect of the cosmic (systematic) variance and
putting it into the error calculation to determine real confidence intervals
around each photometric redshift.

\subsection{Estimates of the photometric error}

We have described in previous works the effect of photometric errors. For a
more complete analysis the reader is referred to LFY98, FLY99, or Y00. As a
resume, we estimate the effects of the photometric error on $z_{\rm ph}$ by
producing fake catalogues of galaxies with given input redshifts ($z_{\rm
in}$) and apparent magnitudes. We assume all of them have the exact SEDs we
use in our technique, hence elliminating the error due to SED
variance. After creating the catalogues and adding to each flux an amount of
noise given by the apparent magnitude, we calculate a photometric redshift
($z_{\rm out}$) for each galaxy. Repeating this calculation a large number of
times for each redshift and apparent magnitude interval, we observe that the
effects of photometric inaccuracies begin to affect the value of the redshift
measurement at $AB(8140)\approx26$. This effect is {\it by definition}
reflected in the redshift likelihood function, which shows a very narrow peak
(width $\Delta z < 0.05$) for bright objects and increasingly wider peaks
(and possibly multiple maxima) for the fainter ones.

\subsection{Estimates of the systematic error}

Whereas in order to estimate the effect of the photometric uncertainties all
we had to do was to eliminate the cosmic variance by simulating spectra in
complete agreement with the model SEDs, now we need to eliminate the
photometric uncertainty in order to estimate the systematic error--the effect
of the cosmic variance.

The path we follow to achieve this is the creation of a large catalogue of
bright objects for which reliable measurements of the redshift are
available. They being bright, we can assume the effect of photometric
uncertainty will be negligible, and the dispersion of the $z_{\rm ph}$ values
around the `real' ones will allow us to estimate the effect of the systematic
error.

\begin{figure}
\label{figzphzsp}
\vbox to 100mm{
\psfig{figure=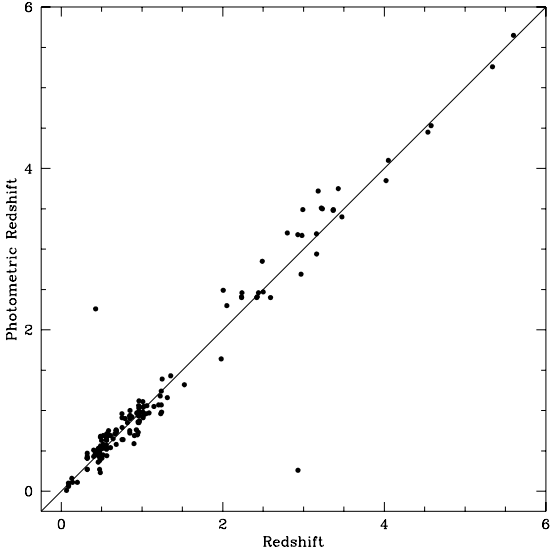,height=80mm}
\caption{Photometric and `real' redshift measurements for the 153 objects in 
our sample. The diagonal line corresponds to $z_{\rm ph}=z_{\rm real}$. Note
the two apparently discordant objects that are not used for the calculation
of $\Sigma$.}
\vfil}
\end{figure}

As was described in Section \ref{seccat}, we will use a catalogue containing
153 sources with reliable spectroscopic redshifts and photometric redshifts
measured by us. Figure \ref{figzphzsp} shows the plot of $z_{\rm ph}$ versus
$z_{\rm real}$\footnote{We use the subindex {\it `real'} in this sample to
indicate that this is not the original spectroscopic list but the one that
has been reanalysed as described above}.

The plot shows that two galaxies show apparent contradiction between the
photometric and spectroscopic values. They are objects \#687 and \#1035 in
FLY99, for which $z_{\rm sp}=$2.931 (C00) and 0.425 (D01) respectively,
compared to $z_{\rm ph}=$0.26 and 2.26. We will not use them in the
calculations described below but will return to them in the next
Section. Apart from those two exceptions, the general agreement is good. We
will now try to model the dispersion of the values using a simple function.

The model we choose to parameterise the systematic error is a normal
distribution with zero mean (we have already proved that the method has no
biases, see FS01) and variable sigma $\sigma_z=\Sigma (1+z)$. This is a very
simple model, whose mathematical form is driven by the fact (proved in FLY99,
see Figure 7 therein) that the value of $\sigma_z/(1+z)$ is approximately
constant over the whole redshift range studied. We would like to remark,
though, that this model is certainly too simplistic--it does not take into
account that at high redshift ($z \gtrsim 3$) the accuracy of the photometric
redshift is expected to improve due to the increased strength of the
Lyman-$\alpha$ break, the decreased variance in the intergalactic {\sc Hi}
absorption, and the reduction of the rest-frame widths of the observed
filters. We feel, however, that the amount of data available at those
redshifts is not enough to warrant an adequate modelization of all those
effects at this stage.

We estimate the value of $\Sigma$ using a maximum likelihood method:

\begin{equation}
L(\Sigma)=\prod_{i=1}^{151} P_i(z_{\rm real}^{(i)};\Sigma)
\end{equation}

\noindent where $P_i(z_{\rm real}^{(i)};\Sigma)$ is the probability of the $i$-th
object being at redshift $z_{\rm real}^{(i)}$ given the value of
$\Sigma$. This probability is calculated by convolving our redshift
likelihood function (which includes the photometric error) with a gaussian of
variable sigma $\sigma_z=\Sigma (1+z)$ (which will account for the systematic
error) and normalising to unit area.

\begin{figure}
\label{figlik}
\vbox to 100mm{
\psfig{figure=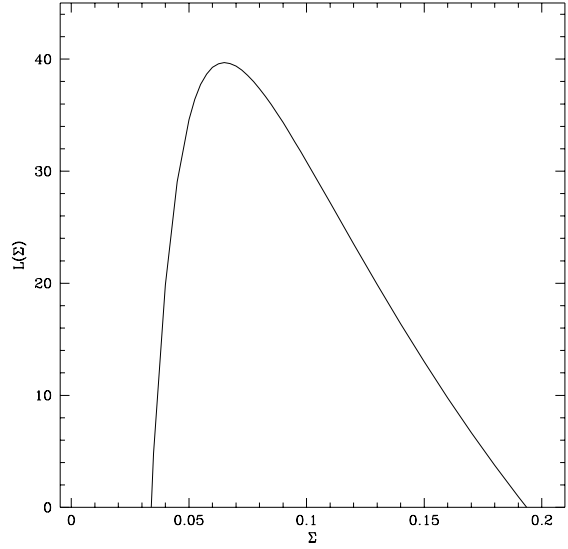,height=80mm}
\caption{Likelihood function for the value of the dispersion parameter
$\Sigma$ given our sample of redshifts and likelihood functions.}
\vfil}
\end{figure}

The result of this likelihood calculation (see Figure \ref{figlik}) is that
we determine the value of $\Sigma$ to be $0.065 \pm 0.003$. In the following
we will use $\Sigma=0.065$, which agrees perfectly with values quoted
previously (but that were calculated only as a measurement of the dispersion
of the photometric measurements).

\section{Calculation of confidence limits}
\label{sectec}

We do now have the tools in hand to perform a careful systematic analysis of
the errors associated to our photometric redshift measurements. For each
single object we have obtained a redshift likelihood function which accounts
for all the photometric uncertainties, and now we also have an estimate of
the systematic uncertainties introduced by the use of a small number of
SEDs--which we assume are independent of the photometric quality ({\it i.e.},
independent of the apparent magnitude of the object).

\begin{figure}
\label{twoobj}
\vbox to 110mm{
\psfig{figure=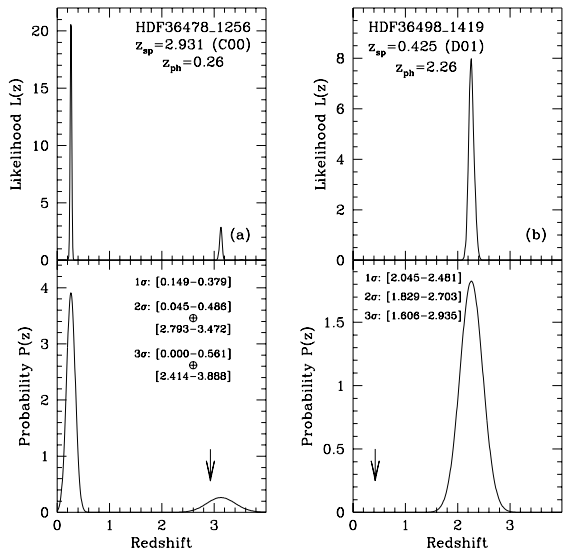,height=80mm}
\caption{Likelihood (top) and probability (bottom) functions for the two
objects with apparently discordant redshifts. Panel (a) shows the results for
HDF36478\_1256 and panel (b) for HDF36498\_1419. The vertical arrows in the
lower panels mark the spectroscopic value.}
\vfil}
\end{figure}

How do we combine these two things? We choose to obtain the {\it probability
function} $P_i(z)$ for the $i$-th object as the convolution of its redshift
likelihood function $L_i(z)$ (normalised to unit area) with the gaussian of
variable sigma described above:

\begin{equation}
P_i(z) = \int_{z'=0}^{z'=\infty}dz'\ L_i(z')\ G\left[z|z',\Sigma(1+z')\right]
\end{equation}

\noindent where $G$ is a gaussian distribution of median $z'$ and 
$\sigma_z=\Sigma(1+z')$, truncated at $z<0$ and normalised to unit area. As an
example of the calculation of this probability, we show in Figure
\ref{twoobj} the likelihood and probability functions for the two objects
that show apparently discordant values of the photometric and spectroscopic
redshifts.

Once the probabilities $P(z)$ are obtained, it is trivial to define
confidence intervals for the value of $z_{\rm ph}$. We choose to do it by
defining the confidence interval at probability $p$ as the region $\cal Z$ in
redshift space such that (i) $P(z)>l \: \forall z \in \cal Z$ and (ii) $\int_{\cal
Z} P(z) \,dz = p$, with $l$ being the value of $P(z)$ at the limits of the
region $\cal Z$. Observe that the region $\cal Z$ ({\it i.e.}, the confidence
interval) need not be connected, as will generally happen in those cases 
where $P(z)$ shows multiple peaks.

\begin{figure}
\label{conflim}
\vbox to 110mm{
\psfig{figure=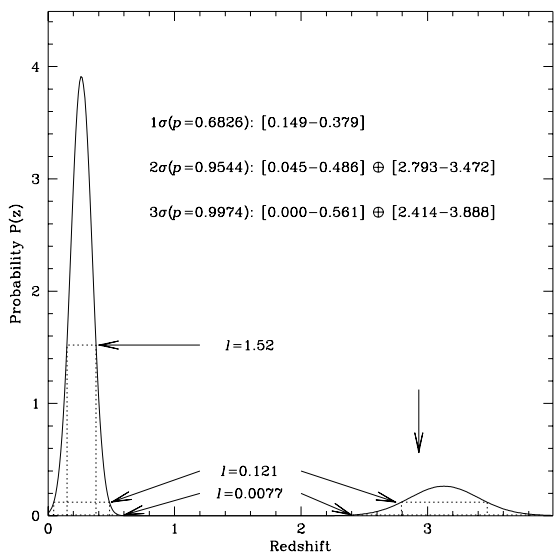,height=80mm}
\caption{Calculation of the confidence limits for object HDF36478\_1256. The
vertical dotted lines mark the redshift ranges that enclose 0.6826, 0.9544,
and 0.9974 of the total probability. Notice that the latter two are disjoint
regions, and that the last one is almost invisible because of its proximity
to the X axis. As in Figure 3, the vertical arrow marks the position of the
spectroscopic value.}
\vfil}
\end{figure}

This slightly abstruse definition, in practice, comes down to finding the
points where a horizontal line cuts $P(z)$ such that the area inside the
curve at the cut points is equal to the value of $p$. For convenience we
define the $1\sigma$ region to be that for which $p=0.6826$, $2\sigma$ at
$p=0.9544$, and $3\sigma$ at $p=0.9974$. In Figure \ref{conflim} we show this
process in detail for object HDF36478\_1256.

\section{Discussion}
\label{secdis}

\begin{figure}
\label{zszplim}
\vbox to 100mm{
\psfig{figure=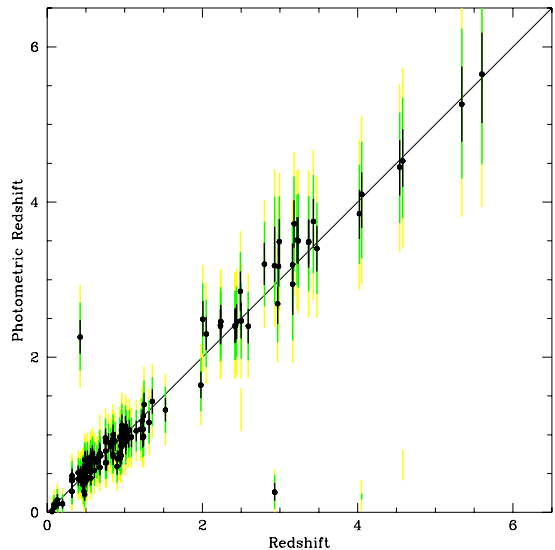,height=80mm}
\caption{Plot of $z_{\rm ph}$ versus $z_{\rm real}$ including the 1-, 2-, and
3-$sigma$ confidence intervals for each photometric redshift (drawn as
increasingly pale shades of grey). Notice the presence of `isolated' low-$z$
pieces of confidence intervals belonging to high redshift objects.}
\vfil}
\end{figure}

We can now calculate confidence intervals for all objects in our catalogue,
in particular for the 153 objects which have secure spectroscopic redshifts.
Figure \ref{zszplim} is an attempt to show all 153 points together with their
confidence limits. This is made difficult by the crowding of objects.

\subsection{The catastrophic errors}

The term ``catastrophic error'' has been widely used in the photometric
redshift bibliography to identify those cases where the photometric and
spectroscopic measurements of the redshift differ in an amount much larger
than the expected systematic dispersion (usually at a level greater than
three or four times the average photometric redshift dispersion). It has been
usually assumed that these cases highlight objects for which the photometric
technique fails, possibly due to the lack of sufficiently detailed spectral
models to reproduce the intrinsic source spectrum.  However, our experience
proves (see FS01) that in a large majority of these situations, the
spectroscopic value was the one that needed revision.

Nevertheless we know that the redshift probability functions, as described
above, can in some cases be multimodal. In such cases, even if our technique
is working properly, it may yield results for the best-fit redshift that are
distant from the exact value. These events would be considered ``catastrophic
errors'' using the traditional meaning of the term (and also in a purely
scientific sense, because they are product of a bifurcation in parameter
space). We prefer not to call them ``catastrophic'', given that the
results--once the error bar is considered--are in fact not in error, and that
they can be perfectly individuated using the techniques described in this
paper.

Let us study the two particular cases that appeared in Figure \ref{twoobj} as
objects with apparently discordant redshifts. It can be seen that in the case
of object HDF36478\_1256 the photometric and spectroscopic redshifts are in
fact compatible to within a $2\sigma$ confidence level. This case was already
discussed in FS01, and the conclusion is that it cannot be considered an
error (even less a {\it catastrophic} error) as the real value is perfectly
within the confidence limits of our measurement.

The case of HDF36498\_1419 is different. Our redshift probability curve is
absolutely incompatible with the redshift claimed in D01. The authors have
kindly supplied us with the spectrum, which shows an obvious emission line at
$\lambda = 5311$ \AA. The identification of this line with {\sc [Oiii]}
$\lambda3727$ is not evident because of the lack of, amongst others, {\sc
H}$\beta$ and {\sc [Oiii]}$\lambda\lambda$4959,5007 in the covered wavelength
range--the authors consider this redshift determination only `tentative'. It
is true however that other identifications that would put the redshift in
agreement with the photometric measurement ({\sc Ciii]} $\lambda1909$ or {\sc
Civ} $\lambda\lambda$1548,1551) are also problematic because of the absence
of other important lines in the observed range.

We have checked our photometry and do not see any particular problem with the
object--it is isolated, and it does not look likely that light from nearby
objects could either fool our photometry or produce spurious emission
lines. It is not clear what is causing the divergence.

\subsection{A check of the confidence intervals}

In order to check that the confidence intervals we are calculating are
consistent, we perform the following test: for each object, we observe
whether the spectroscopic (`real') redshift is consistent with our value to
within a $1\sigma$, $2\sigma$, $3\sigma$ or none of them. Given the
uncertainty about HDF36498\_1419 discussed in the previous paragraph, we do
not include it in this calculation. The results are listed in Table 2,
together with a comparison with the expectations for a pure normal
distribution.

\begin{table}
 \centering
  \caption{Analysis of the confidence limits. The columns list the number and
fraction of objects whose real redshift is within the 1, 2, and 3 $\sigma$ 
confidence intervals from our measurement. For comparison, the fractions
expected from a normal distribution are also tabulated.}
  \begin{tabular}{l c c c}
 Confidence interval & Obs. number & Obs. fraction & Normal \\
 $<1\sigma$            & 105/152     & $0.691 \pm 0.067$   & 0.6826 \\
 $<2\sigma$            & 145/152     & $0.954 \pm 0.079$   & 0.9544 \\
 $<3\sigma$            & 151/152     & $0.993 \pm 0.081$   & 0.9974 \\
 $>3\sigma$            &  1/152      & $0.007 \pm 0.007$   & 0.0026 
\end{tabular}
\end{table}

It must be noticed that the object which is further than $3\sigma$ is
actually only marginally so: the spectroscopic redshift is $z_{\rm sp}=0.483$,
with the three sigma interval around $z_{\rm ph}=0.230$ being
[0.000--0.473]. In any case, the presence of a $>3\sigma$ deviation in
a sample of 151 members is, as indicated in the table, not particularly
remarkable. 

The results in Table 2 are indicative of the accuracy of our error
estimates. We think that this proves that the method here described is a
consistent and efficient way to estimate the errors associated to the
photometric redshift measurements, and that this method should be used in
order to obtain error estimates of any quantity which is measured based on
catalogues of photometric redshifts.

\section{Conclusions}
\label{seccon}

We have presented a complete analysis of the sources of error present in the
photometric redshift determination technique. After showing in a previous
paper (FS01) that photometric redshifts are at least as reliable as
traditional spectroscopic redshifts when it comes to the measurement of faint
galaxies, we show in this paper that the photometric redshifts have an
additional advantage: the error in the measurement can be completely
characterised by means of a redshift probability function.

We have described how this probability function can be obtained for each
individual object, making use of the redshift likelihood function (which
accounts for the photometric uncertainties) and of the error component
that is added as a systematic effect by our technique. We have modelled this
component as a gaussian with variable variance, $\sigma_z=\Sigma (1+z)$, and
measured the parameter $\Sigma=0.065$ from a large sample of reliable
redshifts. 

By convolving both error components we obtain a {\it redshift probability
function} for each object, that easily allows for the determination of
confidence intervals associated to the value of the photometric redshift.

We have checked that the confidence limits thus calculated are consistent,
and suggest that any quantity to be measured from photometric redshift
catalogues in the future should make use of similar techniques, in order to
account for the errors inherent to the process of redshift determination.

\section{Acknowledgments}

We would like to thank our referee, Mark Lacy, for his insightful comments
that have improved the clarity of the paper. AFS gratefully acknowledges the
support of the European Commission through a Marie Curie Fellowship.

\label{lastpage}

\end{document}